\documentclass[twocolumn,prb,showpacs]{revtex4}
\usepackage{psfrag}
\usepackage{graphicx}
\usepackage{amsmath}
\usepackage{amssymb}

\usepackage{color}

\newcommand{\ud}{\mathrm{d}}

\begin{document}
\title{Controlling the transmission line shape of molecular t-stubs and potential thermoelectric applications}
\author{Robert Stadler$^1$ and Troels Markussen$^2$}
\affiliation{$^1$Department of Physical Chemistry, University of Vienna, Sensengasse 8/7, A-1090 Vienna, Austria\\
$^2$Center for Atomic-scale Materials Design (CAMD), Department of Physics, Technical University of Denmark, DK-2800 Kgs. Lyngby, Denmark} 
\date{\today}

\begin{abstract}
Asymmetric line shapes can occur in the transmission function describing
electron transport in the vicinity of a minimum caused by quantum interference
effects. Such asymmetry can be used to increase the thermoelectric efficiency
of molecular junctions. So far, however, asymmetric line shapes have
been only empirically found for just a few rather complex organic molecules where the origins of the line shapes relation to molecular structure were not resolved. In the present
work we introduce a method to analyze the structure dependence of the asymmetry of
interference dips from simple two site tight-binding models, where one site corresponds to a molecular $\pi$ orbital of the wire and the other to an atomic $p_z$ orbital of a side group, which allows us to analytically characterize the peak shape in terms of just two parameters.
We assess our 
scheme with first-principles electron transport calculations for a variety of {\it t-stub} molecules and also address their suitability for thermoelectric applications.
\end{abstract}
\maketitle

\section{Introduction}

Despite increased interest in molecular electronics and tremendous advances in the theoretical description as well as experimental characterization of the conductance of single molecule junctions in the last two decades, there has been a shortage of novel ideas for molecular devices. Recently, two concepts emerged which received a lot of attention within the field: 
i) devices based on quantum interference (QI) effects~\cite{Baer2002,Stadler2003,StadlerNanotech2004,DijkOrgLett2006,Andrews2008,MarkussenJCP2010,Kocherzhenko2011} and ii) theromelectric applications (TA)~\cite{Paulsson2003,Ke2009,Bergfield2009,Finch2009,Dubi2009,Bergfield2010,Nozaki2010,Leijnse2010,Quek2011,Dubi2011,KamalMarkussen2011}. 
QI devices are utilizing the wave nature of electrons and the resulting possibility for destructive interference of transmission amplitudes for the design of data storage elements~\cite{Stadler2003}, transistors~\cite{Baer2002,Andrews2008,Kocherzhenko2011} or logic gates~\cite{StadlerNanotech2004}, and QI is also the cause for the observed low conductance of a benzene contacted in the meta configuration as compared to the para and ortho configurations~\cite{Patoux1997,Mayor2003}. It has been rationalized by a variety of different physical pictures, such as phase shifts of transmission channels or interfering spatial pathways\cite{Sautet1988,Yoshizawa2008,Fowler2009,Hansen2009}. More recently, research in QI has been extended to aromatic molecules of increasing
size~\cite{Walter2004,Cardamone2006,Stafford2007,Ke2008} as well as incoherent transport in the Coulomb blockade regime~\cite{Hettler2003,Donarini2009} and also led to proposals for the usage of molecular interferometers in spintronics~\cite{Hod2006,Cohen2007}.

Thermoelectric materials on the other hand can convert thermal gradients to electric fields for power generation or vice versa for cooling or heating which makes them useful as Peltier elements. The efficiency of a thermoelectric material is measured by a dimensionless number, the figure of merit $ZT=S^2TG/\kappa$, which includes the thermopower $S$, the average temperature $T$, the electronic conductance $G$, and the total thermal conductance $\kappa$, which has contributions from electrons ($\kappa_{el}$) as well as from phonons ($\kappa_{ph}$). In Appendix A we show how these quantities are calculated. Materials with $ZT\sim1$ are regarded as good thermoelectrics, but $ZT>$ 3 would be required to compete with conventional refrigerators or generators~\cite{Majumdar2004}. The optimal thermoelectric material has a high thermopower and high electronic conductance but a low thermal conductance. One way to optimize $ZT$ is thus to increase the thermopower. It is instructive to consider an approximate formula for the thermopower~\cite{Paulsson2003} 
\begin{equation}
S=\left.\frac{T(\pi k_B)^2}{3e}\frac{d\ln(\mathcal{T}(E))}{dE}\right|_{E_F},
\end{equation}
where $\mathcal{T}(E)$ is the energy dependent electron transmission function.  This formula is valid at low temperatures and at energies not too close to transmission resonances. By inspection of Eqn. (1) it becomes obvious that a large thermopower can be achieved if the transmission function changes rapidly around the Fermi energy. This property therefore provides a link between QI and TA in the sense that QI effects can create steep declines in the transmission function for certain molecules. Although quite good thermo-electric characteristics have also been predicted for some systems with the QI induced minimum surrounded by two maxima symmetrically~\cite{Bergfield2010}, ideally the minimum is closer in energy to one molecular resonance where the transmission has a maximum thus creating a very asymmetric transmission function, which naturally results in a steep transmission gradient in the respective energy range. 
In studies of mesoscopic quantum waveguides such asymmetric peak shapes caused by QI are usually referred to as Fano resonances,\cite{Papadopoulos2006,Kormanyos2009} which are not limited to electron transport but a quite general phenomenon in  physics.\cite{Miroshnichenko2010} They occur due to the interference between two wave amplitudes, one corresponding to a continuous state and the second to a localized state, and were originally observed in the line-shapes of inelastic electron scattering by helium atoms, where the continuous state was defined by the background scattering process and the localized one by the process of autoionization~\cite{Fano1961,Fano1965}. 

\begin{figure}
\includegraphics[width=0.9\linewidth,angle=0]{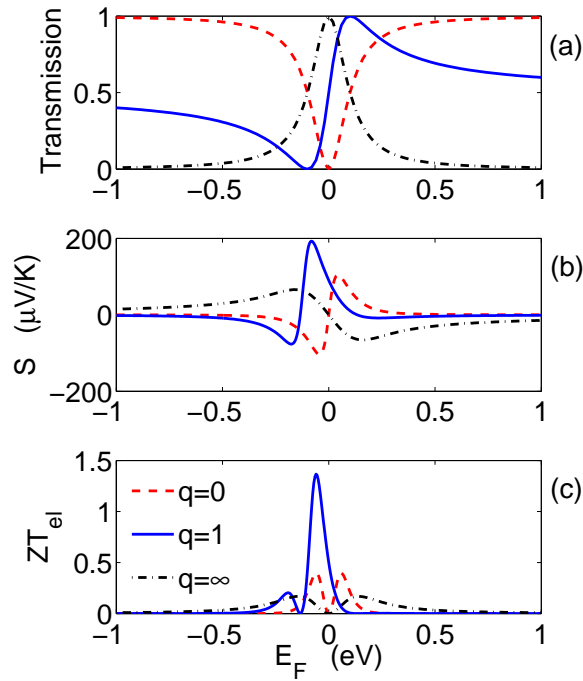}
\caption[cap.Pt6]{\label{fig.FanoShapes} (a) Transmission function, (b) thermopower, and (c) purely electronic contribution to the thermoelectric figure of merit ZT$_{el}$ (i.e. the phonon contribution to the thermal conductance is neglected). The different curves are calculated from the general Fano-transmission function giving a symmetric dip for $q=0$ (dashed red), an asymmetric Fano-shape for $q=1$ (solid blue), and a symmetric peak for $q\rightarrow \infty$ (dash-dotted black). The temperature is $T=300\,$K and the broadening parameter is $\Gamma=0.2\,$eV.}
\end{figure}

As an illustrative example of the effect of transmission function asymmetry on $S$ and $ZT$, we consider a model system with a transmission function described by the Fano function
\begin{equation}
\mathcal{T}(E)=\frac{(q\Gamma/2+E)^2}{(1+q^2)(E^2+\Gamma^2/4)}. \label{eq:Fano-def}
\end{equation}
The defining feature of a Fano resonance is that the resonance profile of the scattering cross section can be brought into this form. In Fig. \ref{fig.FanoShapes} (a) we plot the transmission function $\mathcal{T}(E)$ in Eqn.~\eqref{eq:Fano-def} for three different values of the shape parameter $q$ giving rise to a symmetric dip ($q=0$), an asymmetric Fano-shape ($q=1$) and a symmetric transmission peak ($q\rightarrow\infty$). Panels (b) and (c) in Fig. \ref{fig.FanoShapes} show the corresponding thermopower and $ZT$, respectively. Clearly the asymmetric transmission function gives the largest thermopower and $ZT$. At lower temperature and smaller $\Gamma$-values, the transmission peak will result in significantly higher $S$ and $ZT$, but the asymmetric Fano-shape will always give larger values than the symmetric transmission dip.  

Although the antiresonances found in mesoscopic electron waveguides with stub-tuners~\cite{Sols1989,Porod1992,Porod1993,Debray2000} can be classified as Fano resonances, because there the direct transmission supported by a continuum of states within a periodically extended wire interferes with the states localized in the stub, the analogy does not hold for single molecule junctions where the continuum of the electrode states is interrupted by the junction which acts as a tunnelling barrier~\cite{Emberly1999}. But even if the analogy is reduced to a similarity, asymmetric QI features (which at least resemble Fano resonances) have been observed also in the transmission functions describing electron transport through single molecules~\cite{Papadopoulos2006,Stadler2009}. Moreover, the general principle that electron waves propagating from the left to the right electrode can interfere with localized states in the junction can also apply in the coherent tunnelling regime.

What has been lacking so far, however, is a systematic investigation of the dependence of the peak shape of QI induced transmission dips on characteristic features of the structure of the single molecule in the nanojunction, which could match the level of understanding which has been achieved for mesoscopic waveguides. Recently, we established a relation between molecular structure and the occurrence or absence of QI transmission zeros, which could be formulated in terms of simple graphical rules~\cite{MarkussenNanoLett2010,MarkussenPCCP}. In the current article we aim at developing a framework which allows to project the characteristic features of molecular wires with one side group, which are most similar to the mesoscopic stub-tuner, onto simple tight-binding (TB) models. Even for such relatively simple molecules there are structural elements such as the chemical nature of the side group or its exact position on the wire contributing to the peak-shape in a way that is not self-evident. In our article we pursue a step-by-step approach to this problem where we first ask the question how the transmission peak shape and thermoelectric properties are related to the parameters in a two-site tight binding model in general mathematical terms. Then we relate this model to the topology of the $\pi$ states of conjugated organic molecules with side groups and where this process results in chemically stable molecular structures validate our conclusions with first principle electron transport calculations.

\begin{figure}
\includegraphics[width=0.95\linewidth,angle=0]{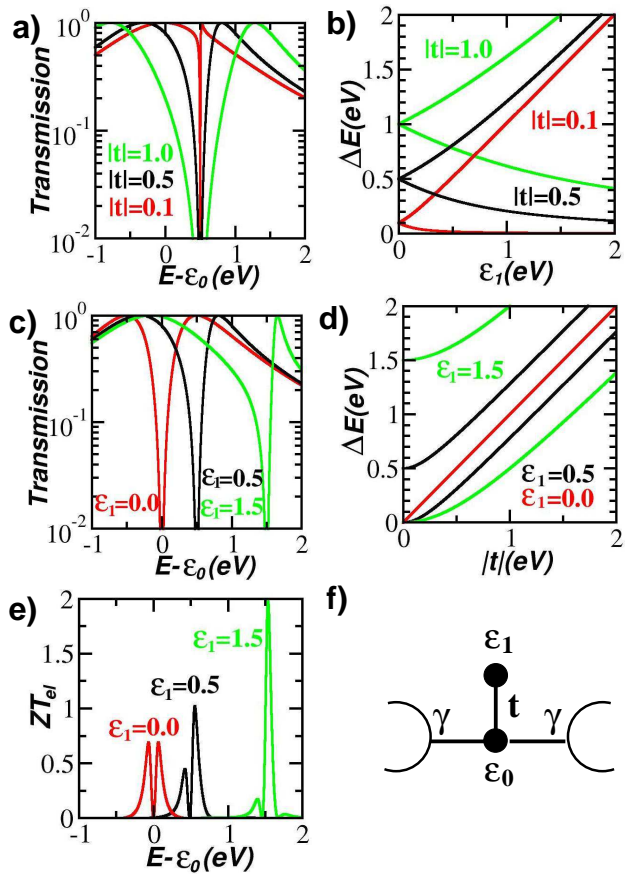}
\caption[cap.Pt6]{\label{fig.TBanalytical}Analytical TB results for a two-site TB model (f) according to Eqns. (1)-(3) with $\gamma = 0.5\,$eV and $\varepsilon_0=0.0\,$eV. (a) Transmission function $T$ as a function of the electron incident energy $E$ (with $\varepsilon_1=0.5\,$eV) and (b) $\Delta E$ as a function of the side group AO energy $\varepsilon_1$ with the MO-AO coupling $|t|=0.1\,$eV (red), 0.5 eV (black) and 1.0 eV (green). (c) $\mathcal{T}(E)$ (with $|t|=0.5\,$eV) and (d) $\Delta E$($|t|$) for $\varepsilon_1=0.0\,$eV (red), $0.5\,$eV (black) and $1.5\,$eV (green). Panel (e) shows the purely electronic thermoelectric figure of merit $ZT_{el}$ calculated from the transmission functions in panel (c). }
\end{figure}

\section{Two-site tight binding model}
The TB-model, which we introduce in the following aims at a separation of the structural aspects governing the peak shape of the interference dips, where analytical expressions can be derived for the transmission function in dependence on just two parameters. For achieving this aim we make a few simplifying assumptions. First, we assume that only the $\pi$ electrons are relevant in the energy range of interest, which allows us to describe the molecules within a TB model with only one $p_z$ atomic orbital (AO) for each carbon- or hetero-atom. In a second step, we look at molecular orbitals (MOs) of the t-stub wires without the side group attached, which can be obtained by diagonalizing the TB-Hamiltonian for a chain of carbon atoms with a certain length and from them pick the single MO which is closest in energy to the $p_z$ AO or fragment orbital of the side group. If we now assume a two-site single-particle picture where the transmission function for electrons incident on a molecular junction with an energy $E$ is dominated by this MO and its interaction with the side group AO, it can be written as
\begin{equation}
\mathcal{T}(E)= \frac{4\gamma^2}{(E-\varepsilon_0-\frac{t^2}{E-\varepsilon_1})^2+4\gamma^2}, \label{Eq:two-site-trans}
\end{equation}
as derived in Appendix B, with $\gamma$ being the lead coupling within a wide band approximation, $\varepsilon_0$ the wire MO energy, $\varepsilon_1$ the side group AO energy and $t$ the coupling between them. For an illustration see Fig.~\ref{fig.TBanalytical} (f).
As explained above the thermoelectric properties of a molecular junction can be optimized if the transmission function rapidly changes from a minimum to a maximum. As a simple measure of how rapidly the transmission function changes we consider the energetic difference $\Delta E$ between the transmission minimum and the closest maximum. It follows from Eq. \eqref{Eq:two-site-trans} that the transmission is zero at energy $E_0=\varepsilon_1$,~\cite{MarkussenPCCP} while for transmission peaks with $T=1$ the quadratic equation $E-\frac{t^2}{E-\Delta\varepsilon}=0$ has to be solved where we made the substitution $\Delta\varepsilon=\varepsilon_1-\varepsilon_0$ and set $\varepsilon_0=0$ without loss of generality for our following arguments. The solutions to this equation $E_{1,2}$ and the norm of their differences with the energy of the transmission zero $E_0$ can then be obtained as
\begin{eqnarray}
E_{1,2}=\frac{\Delta\varepsilon}{2}\pm\frac{1}{2}\sqrt{\Delta\varepsilon^2+4t^2} 
\end{eqnarray}
and
\begin{eqnarray}
|E_{1,2}-E_0|=|-\frac{\Delta\varepsilon}{2}\pm\frac{1}{2} \sqrt{\Delta\varepsilon^2+4t^2}|, 
\end{eqnarray}
where it can be seen that only for finite $\Delta\varepsilon$ there will always be one solution for $|E_{1,2}-E_0|$ which will be closer to zero than the other. In other words only under this condition will the energy difference of the two transmission maxima from the one transmission minimum be asymmetric for the two-site model, while for $\Delta\varepsilon=0$ we get $|E_{1,2}-E_0|=\pm t$ and the peak shape around the minimum in $\mathcal{T}(E)$ will therefore be symmetric. In the following we will refer to the smaller of the two solutions for $|E_{1,2}-E_0|$ as $\Delta E$, which we use as a measure indicative of thermoelectric efficiency where we expect the thermopower to be high for small $\Delta E$. We also find in practice that in most cases a small value for $\Delta E$ indicates a high level of asymmetry in the peak shape around the minimum of the transmission function but this argument cannot be made in general, where a ratio of the two solutions should be used instead. Another conclusion at this point is that $|E_{1,2}-E_0|$ is completely independent of the lead-coupling $\gamma$. This independence of rapid transmission changes from lead-coupling broadenings is a general advantage of utilizing QI since both transmission peaks and transmission minima are defined entirely by the molecular topology. Temperature effects can, however, smear out very sharp features when $k_BT>\Delta E$. 

In Fig.~\ref{fig.TBanalytical} we illustrate explicitly for the two-site model how the shape of the transmission function changes with the size of $t$ (panel (a)) and $\Delta\varepsilon$ (panel (c)), where the peak shapes are found to be the more asymmetric, the larger $\Delta\varepsilon$ is and the smaller $t$. The variation of $\Delta E$ with $\Delta \varepsilon$ and $t$ are shown in panels (b) and (d). We note that these two parameters correspond to the length and height of the tunnelling barrier introduced in Refs.~\cite{Porod1992,Porod1993} for a separation of the sidearm in mesoscopic waveguide t-stubs, where the authors also found that distinctly asymmetric line shapes only start to appear for weak coupling and large barrier heights.

Figure \ref{fig.TBanalytical} (e) shows the purely electronic $ZT_{el}$ (phonon contribution to thermal conductance is neglected) corresponding to the transmission functions in panel (c). Clearly the maximum $ZT_{el}$ values increase when the transmission function becomes more asymmetric, in correspondence with the findings in Fig. \ref{fig.FanoShapes}. While very large $ZT_{el}$ values have been predicted for symmetric transmission functions\cite{Bergfield2010}, the results in Fig. \ref{fig.TBanalytical} (e) show that asymmetric transmission functions are even more promising for a thermoelectric purpose.

\begin{figure*}                                                                                 
\includegraphics[width=0.95\linewidth,angle=0]{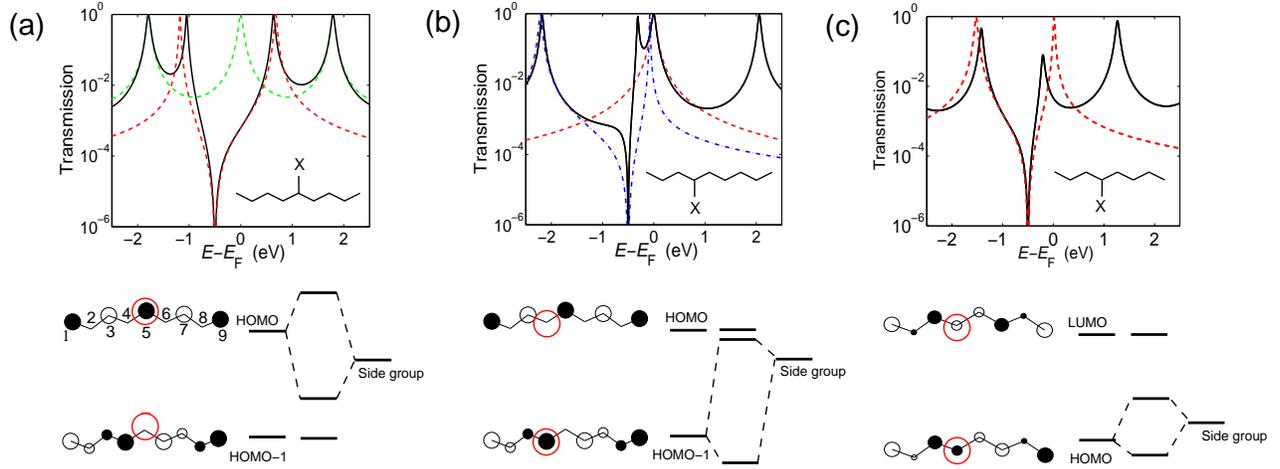}                                     
\caption[cap.Pt6]{\label{fig.scheme}                                                 
Transmission functions (upper panels) and MO-AO hybridisation schemes (lower panels) for (a) C9-5, (b) C9-4 and (c) C8-4 type molecular stubs with a side group AO X ($\varepsilon_{sg}$=-0.5 eV), where the respective AO topologies are given as insets in the transmission figures. The labels 5 and 4 correspond to the atom number in the carbon chain to which the side group is coupled, where we indicate the numbering we use for the atoms in the carbon chain in the lower part of panel (a). The line types for $\mathcal{T}(E)$ correspond to full TB AO calculations with (black) and without (green) the side group, and a 2-site MO model with the side group AO and the HOMO ($\varepsilon_{HOMO}$=0 eV, dashed red) or the HOMO-1 ($\varepsilon_{HOMO-1}$=-1.8 eV, dashed blue), respectively. The Fermi level E$_F$ is assumed to be identical to $\varepsilon_{HOMO}$ of C9 (without any side groups). The lower panels also show the topologies of the HOMO and HOMO-1 where the respective position of the side group is indicated by red empty spheres, where depending on $|\Delta\varepsilon|$ and $|t_{sg}|$ different hybridization patterns are observed.}
\end{figure*}                                                                           

\begin{figure*}
\includegraphics[width=0.95\linewidth,angle=0]{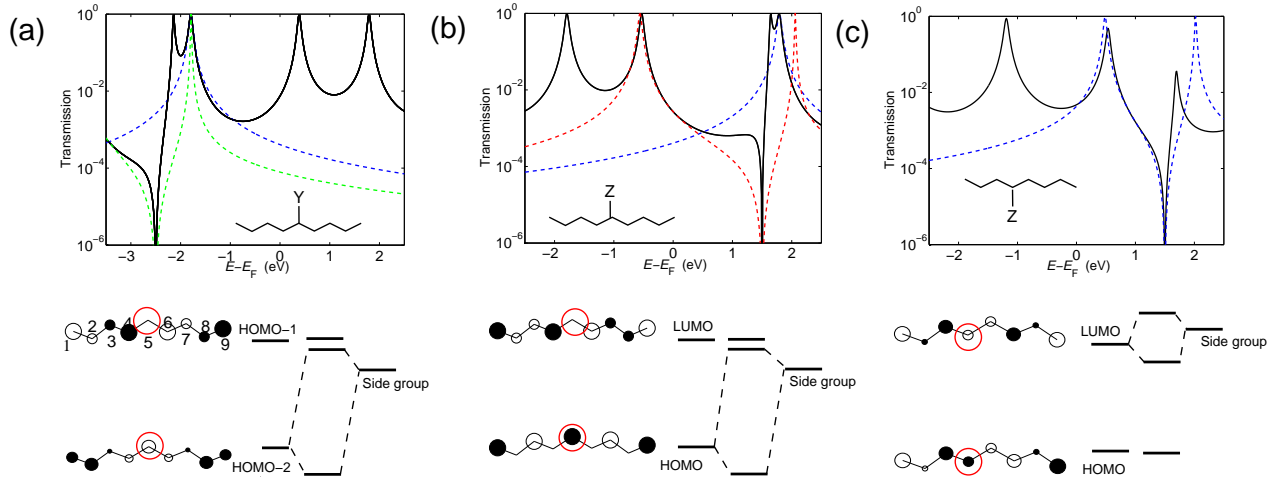} 
\caption[cap.Pt6]{\label{fig.scheme1}Transmission functions (upper panels) and MO-AO hybridisation schemes (lower panels) for the AO topologies (which are shown as insets in the transmission figures) (a) C9-5 with a side group AO Y ($\varepsilon_{sg}$=-2.5 eV), (b) C9-5 and (c) C8-4 with a side group AO Z ($\varepsilon_{sg}$=1.5 eV). The low on-site energy of AO Y is energetically closest to the HOMO-1 (a), but since this does not have any orbital weight on the central carbon atom 5 it does not interact with the side group and produces a Lorentzian transmission function (dashed blue). A reasonable description of the transmission minimum is obtained from the HOMO-2  and the side group (dashed green). In a similar way, AO Z has a side group energy close to the C9 LUMO (b), but since it has no orbital weight on chain site 5 it does not couple to the side group (dashed blue), while the interaction with the HOMO (dashed red) becomes the dominant one with a good qualitative description of the full transmission (solid black). For C8-4-Z (c) the side group is again close to the LUMO, which does have a finite orbital weight on chain site 4 and a 2-site model with the LUMO and the side group (dashed blue) closely reproduces the full TB-AO result (solid black).
}
\end{figure*}   

\section{Molecular orbital analysis}

In order for our two-site model to be useful for the analysis of t-stub molecular wires we need to be able to deduce the parameters $t$ and $\Delta\varepsilon$ from the full topology of the wires $p_z$ AOs including the side group, which does not always have to be another carbon-site. For this we start from an AO-TB-model where the onsite energy of each $p_z$ orbital of the carbon chain $\varepsilon_C$ is set to zero and the couplings between neighbouring sites $\alpha$ are defined as -2.9 eV. From a diagonalization of the chain \textit{without} the side group, where we refer to Appendix C for the full mathematical details, we obtain its MO energies and the orbital weight on each chain site for each MO. Within this scheme we can vary the side group AO energy $\varepsilon_{sg}$ and an initial value for the MO-AO coupling $\beta$ freely so that it reflects the side groups chemical nature. This initial value for $\beta$ is then multiplied by the orbital weight at the chain site to which the side group is bonded in order to derive the coupling parameter $t_{sg}$. The relevant MO is typically the one closest in energy to $\varepsilon_{sg}$ with a finite value for $t_{sg}$. The two site model finally contains this latter MO (with energy $\varepsilon_{MO}$) and the side group AO, where the lead coupling $\gamma$ in the two-site model is derived from the wide band lead coupling $\Gamma$ used in the full AO-TB model multiplied by the squared orbital weight on the terminal sites in the chain. In Figs.~\ref{fig.scheme} and ~\ref{fig.scheme1} we illustrate how this works in practice and which kind of information can be obtained from such an analysis. At this stage we distinguish the chemical nature of the side groups only by their onsite energies and define $\beta=-2.4\,$eV for all the calculations illustrated in the two figures. But even with $\beta$ fixed to a given value, $t_{sg}$ does vary with the MO topologies as discussed below. The side groups in our AO-TB model are always single AOs which for multi-atom groups such as CH$_2$ and NO$_2$ represent fragment orbitals (FOs) where we explain in Appendix D how to obtain from DFT calculations the onsite energy of the FO most relevant to the QI effect.

We want to stress that the AO-TB model we use in this section is only meant to investigate topological effects in a meanfield description. As we point out in the following, some of the topologies we study here would represent radicals or diradicals if they are translated into chemical structures and for a realistic description of their behaviour in a single molecule transport junction a multi-determinant approach would be needed~\cite{Bergfield2011,Geskin2009}. This is, however, not our focus in this article where the AO-TB model is meant to provide a bridge between the 2-site analysis in the previous section and the NEGF-DFT calculations in the next section. For the latter we limit our discussion only to junctions which are chemically stable and well described within a single-determinant approach. 

Fig.~\ref{fig.scheme} focuses on molecular wires containing a side group X with $\varepsilon_{sg}=-0.5\,$eV simulating a CH$_2$ group. This is attached to a chain with nine atoms (in the following referred to as C9) at the fifth (C9-5, Fig.~\ref{fig.scheme}a) and fourth (C9-4, Fig.~\ref{fig.scheme}b) site, respectively, and to a chain with eight atoms (in the following referred to as C8) at the fourth site ((C8-4, Fig.~\ref{fig.scheme}c). In the insets of Fig.~\ref{fig.scheme} we show the molecular topologies, where only C9-5 represents a chemically stable system, while C8-4 would be a radical and C9-4 even a diradical if the side group is linked to the chain by a double bond. Nevertheless it is instructive to compare the three scenarios on a TB level in order to show that while there is no topological limit preventing transmission peak asymmetry for molecular chains with CH$_2$ side groups, in practice such molecules will always have rather symmetric transmission peaks due to a selection criterion defined by chemical stability.

The solid black lines in the transmission functions in the upper panels of Fig.~\ref{fig.scheme} are the results of full TB AO calculations including the side group, while the dashed red and blue lines come from 2-site TB MO models considering only the side group AO and the HOMO (red) or HOMO-1 (blue) of the unsubstituted chain, respectively. For C9-5 the QI feature in $\mathcal{T}(E)$ is quite symmetric and the black line well reproduced by the red line meaning that just considering the HOMO and the side group AO is sufficient for describing the line shape. The green line in Fig.~\ref{fig.scheme}a shows the TB AO result without the side group, where it can be clearly seen that the peak in $\mathcal{T}(E)$ corresponding to the HOMO splits into two when it hybridizes with the side group AO, whereas the other two visible peaks corresponding to the HOMO-1 and LUMO remain unaffected. The situation is quite different for C9-4 (Fig.~\ref{fig.scheme}b), which exhibits a distinctly asymmetric QI feature and the red line does not reproduce the black one at all and instead represents a single site Lorentzian peak at the HOMO energy. The blue line on the other hand shows very good correspondence with the TB AO transmission, which means that the side group AO interacts mainly with the HOMO-1 and not with the HOMO. 

Our findings can be understood in terms of the topologies of the MOs plotted in the lower panels of Fig.~\ref{fig.scheme}, where the HOMO has a high weight on site 5 but none on site 4 and in contrast the HOMO-1 has its largest weight on site 4 and none on site 5. If $\varepsilon_{MO}$ defines the energy of the MO closest to $\varepsilon_{sg}$ with a relatively high coupling parameter $t_{sg}$, these differences in MO topologies also mean that for C9-5 $|\Delta\varepsilon|=|\varepsilon_{sg}-\varepsilon_{MO}|=0.5\,$eV is substantially smaller than the value for C9-4 (1.49 eV), which explains why in the latter case the QI line shape is much more asymmetric according to the general trends shown in Fig.~\ref{fig.TBanalytical}. We note that due to the symmetry properties of the MOs relevant to our analysis C9-3 gives a transmission function similar to that of C9-5, and C9-2 is similar to C9-4, which we checked but do not show here. If the carbon chain has one site less (C8, Fig.~\ref{fig.scheme}c) no such dependency on the position of the side group can be identified simply because the HOMO has a finite localisation on all sites 3-6, and therefore $|\Delta\varepsilon|=0.5\,$eV for all cases. Due to the spread out HOMO localisation pattern, however, $t_{sg}=0.73\,$eV for C8-4, which explains why it shows a somewhat more asymmetric line shape than C9-5 in Fig.~\ref{fig.scheme}a, where $t=-1.07\,$eV. The topologies in Fig.~\ref{fig.scheme} cover all the inequivalent possibilities for wires with nine or eight carbon atoms in the chain with a single CH$_2$ side group. We found that only one of them is chemically stable (C9-5) and this one has a symmetric transmission minimum. Since there is no reason to assume that shorter or longer wires would result in MO topologies markedly different from those investigated, one has to conclude that such t-stub molecules with a CH$_2$ side group are unlikely to exhibit asymmetric transmission shapes in general.

Fig.~\ref{fig.scheme1} now addresses the question what happens if we choose different side group AOs Y and Z with onsite energies $\varepsilon_{sg}=-2.5\,$ and 1.5 eV, respectively for simulating the effect of O and NO$_2$, while we keep the same value for $\beta$ as in the study on the CH$_2$ substituted wires with side group orbital X above. Interestingly, although the position of the minimum in $\mathcal{T}(E)$ is different (since it is always identical to $\varepsilon_{sg}$ for molecular wires with one side group~\cite{MarkussenPCCP}), it can be already said by visual inspection that there is a close correspondence of the peak shape patterns encountered in Fig.~\ref{fig.scheme} and those found in Fig.~\ref{fig.scheme1}. Figs.~\ref{fig.scheme1}a and b, which mimic C9-5-O and C9-5-NO$_2$ wires, respectively, are very similar to the one mimicking the diradical C9-4-CH$_2$ in Fig.~\ref{fig.scheme}b. This is because they exhibit again quite high values for $|\Delta\varepsilon|$, namely 0.91 (with respect to the HOMO-2) for C9-5-Y and 1.5 with respect to the HOMO for C9-5-Z. C8-4-Z (Fig.~\ref{fig.scheme1}c) on the other hand resembles C8-4-X (Fig.~\ref{fig.scheme}c). In both cases $|\Delta\varepsilon|=0.5\,$ and $t_{sg}=0.73\,$eV, although the relevant MO is the HOMO in Fig.~\ref{fig.scheme}c and the LUMO in Fig.~\ref{fig.scheme1}c.

\begin{figure}
\includegraphics[width=\linewidth,angle=0]{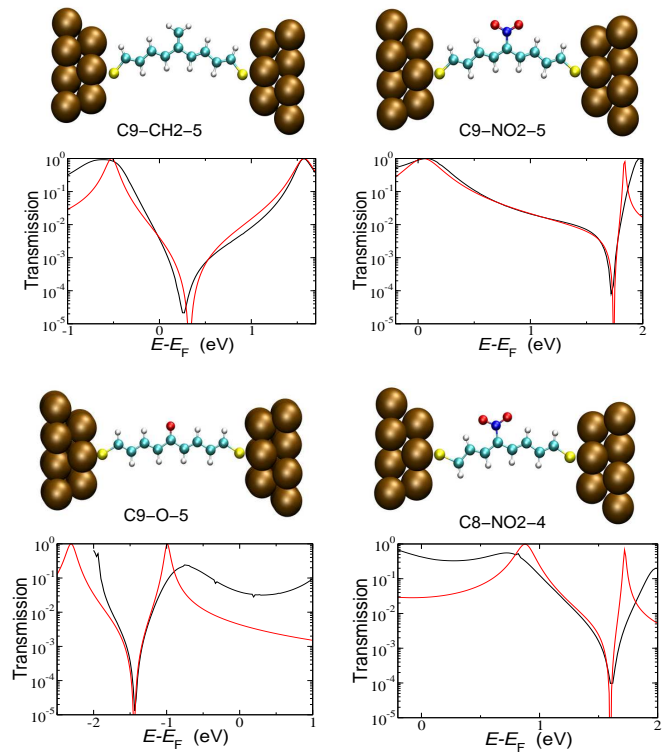}
\caption[cap.Pt6]{\label{fig.dftcurves}Transmission functions calculated for the molecular junctions shown in the inset from full NEGF-DFT calculations~\cite{note-on-DFT} (black lines), and from 2-site MO models (red lines) with the parameters $|\Delta\varepsilon|$ and $|t_{sg}|$ as listed in Table~\ref{tab.dftresults} and obtained directly from DFT. For details see text.}
\end{figure}

\section{First-principles electron transport calculations}
As a next step we test our TB predictions and their interpretation on the basis of 2-site MO models by performing DFT based electron transport calculations~\cite{note-on-DFT} for the transmission function of the four molecules shown in the insets of Fig.~\ref{fig.dftcurves}. As discussed in the previous section some topologies cannot be realised chemically, since they would be chemically unstable. The molecules C9-CH$_2$-5, C9-O-5 and C8-NO$_2$-4 in Fig.~\ref{fig.dftcurves} on the other hand are legitimate targets for organic synthesis. While this is not true for the radical C9-NO$_2$-5, there is still some justification for studying the respective junction with a DFT approach, since the nitro group can stabilize radicals at least as transition states in conjugated $\pi$ systems via the mesomeric effect. We also derive the TB parameters for projecting the transport characteristics of these molecules onto 2-site models from the same calculations. Our procedure for the latter has the following steps: i) The local eigenstates on the side group orbital is found from a subdiagonalization of part of the full Hamiltonian corresponding to the LCAO basis orbitals on the side group atom. This yields information of the side group energy and the coupling to the nearest carbon $p_z$ orbital in the chain (see Appendices C and D for more details).  
ii) For projecting the topology of the molecular $\pi$ states without the side group we use the same scheme as with the AO-TB-models above, where the lead coupling $\gamma$ now is used as a fitting parameter in order to ensure that the broadening in the transmission functions calculated from the 2 site-MO model resembles the DFT transmission. iii) Finally we allow for a rigid energy shift of the 2 site-MO transmission function in order to account for Fermi level alignment in the DFT calculations~\cite{Stadler06,Stadler07,Stadler10}. We show $\mathcal{T}(E)$ for all four molecules as calculated from NEGF-DFT (black lines) and from the corresponding 2 site-MO model calculations (red lines) in Fig.~\ref{fig.dftcurves} and list the obtained model parameters $|\Delta\varepsilon|$ and $|t_{sg}|$ as well as the values for $E_0$ and $|\Delta E|$ obtained from the black curves in Table~\ref{tab.dftresults}.

\begin{table}[tp]
\begin{tabular}{|l||l|l|l|l|}
\hline \hline
  molecule       & C9-CH$_2$-5 & C9-O-4 & C8-NO$_2$-4 & C9-NO$_2$-5 \\
  \hline \hline
$|\Delta\varepsilon|$ & 0.40 &  0.91 & 0.61 & 1.60 \\
$|t_{sg}|$  & 1.03 & 1.12 & 0.32 & 0.40 \\
\hline \hline
$E_0$  & 0.26 &  -1.44 & 1.61 & 1.72 \\
$|\Delta E|$ & 0.89 & 0.65 & 0.40 & 0.20 \\
\hline \hline
$S$ ($E$) & -170 (0.21) & 184 (-1.39) & 199 (1.65) & 349 (1.74) \\
$ZT_{el}$ ($E$) & 0.82 (0.19) & 0.93 (-1.37) & 1.00 (1.67) & 3.00 (1.77) \\
$ZT$ ($E$) & 0.003 (0.19) & 0.01 (-1.37) & 0.03 (1.67) & 0.48 (1.77)\\
\hline \hline
\end{tabular}
\caption{2-site MO model parameters $|\Delta\varepsilon|$ and $|t_{sg}|$ used for the red lines in Fig.~\ref{fig.dftcurves}. The energy of the transmission zero $E_0$ and its difference to that of the the nearest transmission maximum $|\Delta E|$ are evaluated from the DFT transmission curves (black lines in Fig.~\ref{fig.dftcurves}). The maximum values for the thermopower $S$ and the electronic contribution to the figure of merit $ZT_{el}(E)$ (obtained at the corresponding energies in parantheses) and the full figure of merit $ZT$ (including phonon contributions and calculated at the same energy) are listed in the bottom rows. All energies and the coupling strength $|t_{sg}|$ are given in eV, while S is defined in $\mu V/K$, and $ZT_{el}$ and $ZT$ are dimensionless and calculated at the temperature $T=300\,$K.}
\label{tab.dftresults}
\end{table}

The peak shapes in Fig.~\ref{fig.dftcurves} which are now based on DFT calculations are very similar to those obtained from the TB models in Figs.~\ref{fig.scheme} and ~\ref{fig.scheme1}. Now, however, the hopping element $\beta$ has realistic values of -2.4, -2.5 and -1.5 eV for CH$_2$, O and NO$_2$, respectively, which naturally has an effect on $|t_{sg}|$ as listed in Table~\ref{tab.dftresults}. The smallest value for $|\Delta E|$ is found for C9-NO$_2$-5 with the largest $|\Delta\varepsilon|$ and a rather small $|t_{sg}|$, while C9-CH$_2$-5 has the largest $|\Delta E|$ with the smallest $|\Delta\varepsilon|$ and a large $|t_{sg}|$. In both cases the side group couples mostly to the HOMO of the unsubstituted molecule. But because of the difference in the onsite energies of the side group, $|\Delta\varepsilon|$ differs by 1.2 eV between the two, while the weaker coupling of NO$_2$ to the conjugated $\pi$ electrons of the wire is reflected in a reduction of a factor of two in $|t_{sg}|$. Both effects combine to explain the factor of four in $|\Delta E|$ when these two systems are compared. C9-O-5 and C8-NO$_2$-4, where the side group couples strongest to the HOMO-2 and the LUMO, respectively, are in between these two extremes in terms of $|\Delta E|$, where the low coupling parameter $|t_{sg}|$ for the nitro group seems to dominate over the larger value of $|\Delta\varepsilon|$ found for the oxygen side group. 

Finally we want to address the suitability of the four molecules in Fig.~\ref{fig.dftcurves} for thermoelectric devices, where we calculate the thermopower $S$ and the figure of merit $ZT$ explicitly from DFT (for a description of the theoretical framework we use for that we refer to Appendix A and Refs.~\cite{Bergfield2009,Finch2009,Bergfield2010,Nozaki2010}). We calculate both the purely electronic $ZT_{el}$, which is completely determined by the electron transmission function assuming a vanishing phonon thermal conductance, $\kappa_{ph}$, and the full $ZT$ with a constant value of $\kappa_{ph}=50\,$pW/K as it was determined experimentally for quite similar alkanes with thiol anchors~\cite{WangScience2007}.  
 
Table ~\ref{tab.dftresults} lists the calculated thermoelectric properties for the four molecules in Fig.~\ref{fig.dftcurves}. Notably the qualitative ranking in $S$ and $ZT$ reversely corresponds to that of $|\Delta E|$, reflecting that the more asymmetric $\mathcal{T}(E)$ curves result in higher values for thermopower and figure of merit. This remains true and is at the highest end even emphasized when the contributions of phonons are considered. The two drawbacks for C9-NO$_2$-5 in terms of applications are the rather high value of $E_0$ and its instability as a radical. These results demonstrate that for an efficient and robust way of implementing QI induced transmission minima for thermoelectric applications an asymmetric transmission function with a transmission peak located close to the minimum in a chemically stable structure is needed. Since our analysis covers a large part of the chemical possibilities for t-stub type molecules, i.e. simple wires with one side group, one has to conclude that more complex probably aromatic molecules need to be investigated systematically with regard to their suitability for thermoelectric devices.

\section{Summary and Conclusions}
In summary, we presented a scheme for the systematic investigation of the dependence of the symmetry of the transmission function around the energy of quantum interference induced minima on the molecular structure in single molecule junctions. For this purpose we mapped the molecular orbitals of molecular wires with side groups onto simple two-site tight binding models where it was found that the shape of the transmission function just depends on two parameters which provides a strong link to previous findings for t-stub tuners in mesoscopic waveguides. We applied the mapping procedure on a series of molecular wires which vary in their number of carbon atoms as well as the chemical nature and bonding position of side groups. We predicted from our scheme and verified with density functional theory calculations that the peak shape is mostly determined by the coupling between the side group and the molecular chain (where a small coupling results in high asymmetry) and the energy difference of the side group atomic orbital and the molecular orbital closest to it (where a large difference results in high asymmetry). The role of the topology of this molecular orbital in producing quite different peak shapes for very similar molecules could be explained from our model and was further verified by DFT transport calculations. Asymmetric peak shapes are highly desirable in terms of device applications, because they promise high thermoelectric efficiency, and the projection scheme we presented for the systematic study of peak shapes therefore provides a route towards the chemical engineering of thermoelectric devices. 

\begin{acknowledgments}
R.S. is currently supported by the Austrian Science Fund FWF, projects Nr. P20267 and Nr. P22548. The center for Atomic-scale Materials Design (CAMD) receives funding from the Lundbeck Foundation. T.M. acknowledges support from FTP through grant no. 274-08-0408 and from The Danish Center for Scientific Computing. We are deeply indebted to Victor Geskin for helpful discussions.
\end{acknowledgments}

\appendix

\section{Figure of merit - ZT}
The efficiency of a thermoelectric material can be characterized by the dimensionless figure of merit, $ZT$, given by \begin{equation}
ZT = \frac{S^2G\,T}{\kappa_{ph}+\kappa_{el}} \label{ZTdef1},\\
\end{equation}
where $S$ is the Seebeck coefficient, $G$ the electronic conductance, $T$ the temperature, and $\kappa_{ph}$ and $\kappa_{el}$ are the phonon- and electron contributions to the thermal conductance, respectively. Except the phonon contribution to the thermal conductance $\kappa_{ph}$, all quantities can be obtained from the electron transmission function. 

The linear electronic conductance is
\begin{equation}
G(\mu) = \frac{2e^2}{h}\int_{-\infty}^\infty \ud E\,\mathcal{T}(E)\left(-\frac{\partial f(E,\mu)}{\partial E}\right)=e^2L_0 \label{Ge},
\end{equation}
where we have introduced the function $L_m(\mu)$:
\begin{equation}
L_m(\mu) = \frac{2}{h}\int_{-\infty}^\infty \ud E\,\mathcal{T}(E)(E-\mu)^m\left(-\frac{\partial f(E,\mu)}{\partial E}\right) \label{Lm}.
\end{equation}

The thermopower, $S$, is defined as the voltage required to counterbalance the electronic current flowing due to  a (small) temperature gradient across the junction
\begin{equation}
S = -\lim_{\Delta T\rightarrow 0}\frac{\Delta V}{\Delta T}\Big|_{I=0}, \label{S-definition}
\end{equation}
while the purely electronic contribution to the thermopower can be written compactly as~\cite{Sivan1986}
\begin{equation}
S(\mu,T) = \frac{1}{eT}\frac{L_1(\mu)}{L_0(\mu)} \label{S-formula}.
\end{equation}
There is in principle also a so-called phonon drag contribution to the thermopower due to electron-phonon interactions. This term is usually small and we shall not consider it here.

Finally, the electronic contribution to the thermal conductance is given by~\cite{Sivan1986}
\begin{equation}
\kappa_{el}(\mu) =  \frac{1}{T}\left( L_2(\mu)-\frac{(L_1(\mu))^2}{L_0(\mu)} \right)  \label{kappa_e}.
\end{equation}

It is useful to rewrite $ZT$ as
\begin{equation}
ZT = \frac{S^2G\,T}{\kappa_{el}}\cdot\frac{\kappa_{el}}{\kappa_{ph}+\kappa_{el}} = ZT_{el}\frac{\kappa_{el}}{\kappa_{ph}+\kappa_{el}} \label{ZTdef2},
\end{equation}
where $ZT_{el}$ is the purely electronic part which equals the total value for $ZT$, when $\kappa_{el}\gg\kappa_{ph}$. In recent theoretical articles on thermoelectrics in molecular junctions very large values for $ZT_{el}$ have been reported for molecular junctions with quantum interference effects and tranmission nodes~\cite{Finch2009,Bergfield2010}. However, it is not really justified to neglect $\kappa_{ph}$, especially not when the electronic transmission is very small, as is the case around transmission zeros, because in this case $\kappa_{el}$ will also be very small, and the contribution from phonons may be very important.
 
While calculations of the phonon thermal conductance in principle can be performed at a DFT level of theory, the calculations are quite demanding and there are only few reports on first-principles calculations of $\kappa_{ph}$ for molecular junctions~\cite{Nozaki2010,KamalMarkussen2011}. The main focus in our present work is on electronic properties and the $ZT$ values reported in the main text are thus calculated from Eq. \eqref{ZTdef2} assuming a constant value of $\kappa_{ph}$=50 pW/K as determined experimentally for thiol-bonded alkanes~\cite{WangScience2007}, which are quite similar to the molecules considered in our article.

\section{Derivation of the transmission function for two TB orbitals in the wide band limit}

Assuming a single-particle picture, the transmission function for electrons incident on a molecular junction with an energy $E$ can be defined as
\begin{equation}
\mathcal{T}(E) = \text{Tr}[\mathbf{G} \mathbf{\Gamma}_L \mathbf{G}^{\dagger}\mathbf{\Gamma}_R](E)
\end{equation}
where $\mathbf{G}=(E\mathbf{I}-\mathbf{H}_{\text{mol}}-\mathbf{\Sigma}_L-\mathbf{\Sigma}_R)^{-1}$ is the Green's function matrix of the contacted molecule, $\mathbf{I}$ is the identity matrix, $\mathbf{\Sigma}_{L/R}$ is the self-energy due to the left/right lead, and $\mathbf{\Gamma}_{L/R}= i(\mathbf{\Sigma_{L/R}}-\mathbf{\Sigma}_{L/R}^\dagger)$. In our AO-TB models the Hamiltonian describing the molecule is given in terms of a basis consisting of localized atomic-like orbitals, $\phi_1,\ldots,\phi_N$, , where only two are coupled to the leads, which allows us to write the transmission function as
\begin{equation}
\mathcal{T}(E) = 4 \gamma(E)^2|\mathbf{G}_{1N}(E)|^2.
\end{equation}
The matrix element $\mathbf{G}_{1N}(E)$ can be obtained using Cramer's rule
\begin{equation}
\mathbf{G}_{1N}(E)=\frac{C_{1N}(E\mathbf{I}-\mathbf{H}_{\text{mol}})}{\text{det}(E-\mathbf{H}_{\text{mol}}-\mathbf{\Sigma}_L-\mathbf{\Sigma}_R)}
\end{equation}
where $C_{1N}(E\mathbf{I}-\mathbf{H}_{\text{mol}})$ is the $(1N)$ co-factor of $(E\mathbf{I}-\mathbf{H}_{\text{mol}})$ defined as the determinant of the matrix obtained by removing the 1st row and $N$th column from $(E\mathbf{I}-\mathbf{H}_{\text{mol}}-\mathbf{\Sigma}_
L-\mathbf{\Sigma}_R))$ and multiplying it by $(-1)^{1+N}$. Since we assume that only orbitals $\phi_1$ and $\phi_N$ couple to the leads, the removal of the 1st row and $N$th column completely removes $\mathbf{\Sigma}_{L,R}$ in the co-factor.              We imply a wide band limit, where the lead coupling strength, $\gamma$, becomes energy independent and $\mathbf{\Sigma}_{L/R}^\dagger=-i\gamma$. 

For the two-site MO models in the main text there is only one site with an onsite energy $\epsilon_0$ connected to both leads ($N=1$) and coupled to a side group ($\epsilon_1$) with a hopping parameter $t$. For such a system we find

\begin{eqnarray}
C_{11}(E-\mathbf{H}_{\text{mol}})=E-\epsilon_1 & \text{and} & \\ 
\text{det}(E-\mathbf{H}_{\text{mol}}-\mathbf{\Sigma}_L-\mathbf{\Sigma}_R)&=\begin{vmatrix} E-\epsilon_0-2i\gamma & -t  \\
-t & E-\epsilon_1 \\                 \end{vmatrix} \label{2siteH},
\end{eqnarray}
which results in
\begin{widetext}
\begin{equation}
\mathcal{T}(E)= \frac{(E-\epsilon_1)^2 4 \gamma^2}{((E-\epsilon_0)(E-\epsilon_1)-t^2)^2+4 \gamma^2 (E-\epsilon_1)^2} =\frac{4\gamma^2}{(E-\epsilon_0-\frac{t^2}{E-\epsilon_1})^2+4\gamma^2}.
\end{equation}
\end{widetext}

\section{Derivation of the MO projection scheme}
In order to address the explicit dependence of the side group on-site energies and coupling parameters we first write the full Hamiltonian of the molecule (in a basis of $p_z$ orbitals) as
\begin{equation}
H = H_0 + H_{sg},
\end{equation}
where $H_0$ describes the linear carbon chain without the side group having the structure
\begin{equation}
H_{0} = \left(  
\begin{array}{lllll|l}
 \varepsilon_c & \alpha &  & &  & 0 \\ 
 \alpha & \varepsilon_c & \alpha  &  &  & 0 \\ 
 & \alpha &\ddots   & \ddots &  & \vdots \\
 & &\ddots  &  &  &   \\ 
   &        &  &  &  &  \\ \hline
 0 & 0 & \hdots &  & & 0  \end{array}
              \right). \label{H0}
\end{equation}

Here $\varepsilon_c$ is the carbon on-site energy and $\alpha$ is the nearest neighbour hopping parameter. The last row and column correspond to the single basis function on the side group. The side group Hamiltonian, $H_{sg}$ has the form
\begin{equation}
H_{sg} = \left(  
\begin{array}{lllll|l}
 & &  & &  & 0 \\ 
 & &  &  &  & \vdots \\ 
 & & 0 &  &  & \beta \\
 & &  &  &  &  0 \\ 
   &        &  &  &  & \vdots \\ \hline
 0 & \hdots & \beta & 0 & \hdots & \varepsilon_{sg}
                     \end{array}
              \right).   \label{Hsg}
\end{equation}
We now proceed by diagonalizing $H_0$ to find the molecular orbitals (MOs) of the molecule without the side group. The $i'$th MO has the vector form
\begin{equation}
|\psi^{(i)}\rangle = [c^{(i)}_1,\; c^{(i)}_2,\; \hdots,c^{(i)}_N,\;0],
\end{equation}
assuming there are $N$ atoms in the chain, and where $c^{(i)}_j$ is the coefficient corresponding to the orbital weight on site $j$. The last '0' is at the side group orbital where the MO by definition has no weight. We now make the approximation, to consider only a limited number of MOs and neglect all the others. In the simplest case, we consider only the HOMO, $|\psi^{H}\rangle$, and proceed to construct a \textit{model Hamiltonian} including only the HOMO and the side group orbital. In this case our basis consist of two orbitals:
\begin{eqnarray}
|\phi_1\rangle=|\psi^{H}\rangle &=& [c^{H}_1,\; c^{H}_2,\; \hdots,c^{H}_N,\;0] \\
|\phi_2\rangle=e|\psi^{sg}\rangle &=& [0,\; 0,\; \hdots,0,\;1],
\end{eqnarray} 
where the first one is a MO of the molecule without side group and the last one is the side group orbital. The matrix elements of the model Hamiltonian are given by
\begin{eqnarray}
h_{i,j} = \langle \phi_i|H|\phi_j\rangle = \langle \phi_i|H_0|\phi_j\rangle + \langle\phi_i|H_{sg}|\phi_j\rangle. 
\end{eqnarray}
Since $\langle\phi_1|H_0|\phi_1\rangle=\varepsilon_H$ (the energy of the HOMO), $\langle\phi_2|H_0|\phi_2\rangle=0$, $\langle\phi_1|H_{sg}|\phi_1\rangle=0$, and $\langle\phi_2|H_{sg}|\phi_2\rangle=\varepsilon_{sg}$, we arrive at 
\begin{eqnarray}
h =   \left(  
\begin{array}{cc} 
\varepsilon_H & t_{sg} \\
t_{sg} & \varepsilon_{sg}
\end{array}\right),
\end{eqnarray}
where the coupling parameter $t_{sg}$ is given by
\begin{eqnarray}
t_{sg} = \langle\phi_1|H_0+ H_{sg}|\phi_2\rangle = \langle\phi_1|H_{sg}|\phi_2\rangle = \beta c^{H}_j
\end{eqnarray}
and $c^{H}_j$ is the MO orbital weight on the site, which couples to the side group. We thus observe that the effective coupling, $t_{sg}$, to the side group in the model Hamiltonian depends on the real coupling parameter $\beta$ \textit{and} the orbital weight. This described procedure where only one MO is considered is applied for all the molecules in the main text. The transmission through the model system is calculated in the wide band limit with constant self-energies assuming that only one MO couples to both electrodes. The electrode-self energy matrices thus have the form
\begin{eqnarray}
\Sigma_{L,R} =   \left(  
\begin{array}{cc} 
-i\gamma/2 & 0 \\
0 & 0
\end{array}\right),
\end{eqnarray}
and the transmission function is calculated using the standard non-equilibrium Green's function (NEGF) methods also employed in conjunction with density functional theory (DFT) in the main text. 

The value of $\gamma$ is obtained from the lead coupling $\Gamma$ used in the full AO-TB model as
\begin{eqnarray}
\gamma_L &=& \Gamma_L|c_1^H|^2 \\
\gamma_R &=& \Gamma_R|c_N^H|^2 .
\end{eqnarray}
The reason why the orbital weight is squared is that the lead self-energy, is calculated from a surface Green's function, $g$ as
$\Sigma_{L,R} = V_{L,R}^\dagger gV_{L,R}$, where the effective coupling, $V$, between the leads and the molecule can be calculated for the single HOMO orbital in the same way as its effective coupling to the side group shown above. In all our calculations there are no differences between left and right coupling since all the considered orbitals have the same squared weight on both terminal sites of the chain.

\section{Computation of on-site energies}
We calculate the side group on-site energy from the full Hamiltonian matrix,
$\mathbf{H}$, describing the molecule and the Au electrodes. We project onto the
subspace spanned by the basis functions of the side groups
\[
 \mathbf{h}_{sg} = \mathbf{P}_{sg}\mathbf{H}\mathbf{P}_{sg},
\]
where $\mathbf{P}_{sg}$ has diagonal elements on the indices of the side group basis
functions, and zeros elsewhere and similarly obtain also a side group overlap matrix,
$\mathbf{s}_{sg}$. We then diagonalize $s_{sg}^{-1}\mathbf{h}_{sg}$ to find the
side group energies and eigenstates, where for each side group eigenstate $|\psi_i\rangle$, we calculate a coupling matrix element to the closest $p_z$ orbital (assumed to be on atom $n$) in the main chain as 
\begin{equation}
t_i^n = \langle p_z^{n}| \mathbf{H} |\psi_i\rangle,   
\end{equation} 
with $|p_z^n\rangle$ a vector which is one on the index corresponding to the $p_z$ basis orbital on atom $n$ and zero in all other coordinates. Only the side group eigenstates with $\pi$ symmetry will result in a large coupling $t_i^n$. We therefore include only the $\pi$ state which is closest to the Fermi level in our 2 site MO model and take its energy for the parameter $\varepsilon_{sg}$. 


\end{document}